\documentstyle[11pt,epsfig,epsf]{article}
\def\issue(#1,#2,#3){{\bf #1}, #2 (#3)} 

\def\opcit(#1){ {\em op. cit.}, #1}
\def\etal {\em et al.}

\def\APP(#1,#2,#3){Acta Phys.\ Polon.\ \issue(#1,#2,#3)}
\def\ARNPS(#1,#2,#3){Ann.\ Rev.\ Nucl.\ Part.\ Sci.\ \issue(#1,#2,#3)}
\def\CPC(#1,#2,#3){Comp.\ Phys.\ Comm.\ \issue(#1,#2,#3)}
\def\CIP(#1,#2,#3){Comput.\ Phys.\ \issue(#1,#2,#3)}
\def\EPJC(#1,#2,#3){Eur.\ Phys.\ J.\ C\ \issue(#1,#2,#3)}
\def\EPJD(#1,#2,#3){Eur.\ Phys.\ J. Direct\ C\ \issue(#1,#2,#3)}
\def\IEEETNS(#1,#2,#3){IEEE Trans.\ Nucl.\ Sci.\ \issue(#1,#2,#3)}
\def\IJMP(#1,#2,#3){Int.\ J.\ Mod.\ Phys. \issue(#1,#2,#3)}
\def\JHEP(#1,#2,#3){J.\ High Energy Physics \issue(#1,#2,#3)}
\def\MPL(#1,#2,#3){Mod.\ Phys.\ Lett.\ \issue(#1,#2,#3)}
\def\NP(#1,#2,#3){Nucl.\ Phys.\ \issue(#1,#2,#3)}
\def\NIM(#1,#2,#3){Nucl.\ Instrum.\ Meth.\ \issue(#1,#2,#3)}
\def\PL(#1,#2,#3){Phys.\ Lett.\ \issue(#1,#2,#3)}
\def\PRD(#1,#2,#3){Phys.\ Rev.\ D \issue(#1,#2,#3)}
\def\PRL(#1,#2,#3){Phys.\ Rev.\ Lett.\ \issue(#1,#2,#3)}
\def\SJNP(#1,#2,#3){Sov.\ J. Nucl.\ Phys.\ \issue(#1,#2,#3)}
\def\ZPC(#1,#2,#3){Zeit.\ Phys.\ C \issue(#1,#2,#3)}

\def\bra {\langle}
\def\ket {\rangle}

\def\l {\lambda}

\def\r {\rightarrow}

\def\rnot {R\!\!\!/}
\def\bar {\overline}
\def\bbbar {B^0-\bar{B^0}}

\def\psiks {J/\psi K_S}

\def\be {\begin{equation}}
\def\ee {\end{equation}}
\def\bea {\begin{eqnarray}}
\def\eea {\end{eqnarray}}

\def\bc {\begin{center}}
\def\ec {\end{center}}

\begin{document}
\thispagestyle{empty}
\title{
\rightline  {\small{hep-ph/0307259}}
\rightline  {\small{DO-TH 03/10}}\vspace*{0.5cm}
\bf 
Reevaluating Bounds on Flavor-Changing Neutral Current 
Parameters in R-Parity Conserving and R-Parity
Violating Supersymmetry from \boldmath$\bbbar$\unboldmath ~ Mixing}
\author{ 
{\large\bf Jyoti Prasad Saha}
\thanks{Electronic address: jyotip@juphys.ernet.in}\\
Department of Physics, Jadavpur University, Kolkata 700032, India
\and
{\large\bf Anirban Kundu}
\thanks{Electronic address: Anirban.Kundu@cern.ch}
\thanks{On leave from Department of Physics, Jadavpur University, 
Kolkata 700032, India. Present address: Department of Physics,
University of Calcutta, 92 A.P.C. Road, Kolkata 700009, India.}\\
Universit\"at Dortmund, Institut f\"ur Physik, D-44221 Dortmund, Germany} 

\date{\today}
\maketitle

\begin{abstract}

We perform a systematic reevaluation of the constraints on the
flavor-changing neutral current (FCNC) parameters in R-parity conserving
and R-parity violating supersymmetric models. As a typical process, we
study the constraints coming from the measurements on the $B^0$-$\bar{B^0}$
system on the supersymmetric $\delta^d_{13}$ parameters, as well as on the
products of the $\lambda'$ type R-parity violating couplings. Present data
allows us to put constraints on both the real and the imaginary parts of
the relevant parameters.  

\end{abstract}

\noindent PACS number(s): 
11.30.Hv, 12.60.Jv, 14.40.Nd

\newpage

\section{Introduction}

The answer to whether there is any new physics beyond the Standard Model (SM) 
is probably in the affirmative. One of the most promising candidates for
new physics (NP), and also most studied, is Supersymmetry (SUSY), in both
its R-parity conserving (RPC) and R-parity violating (RPV) incarnations.
Unfortunately for the believers of a Theory of Everything, SUSY introduces
a plethora of new particles, and even in its most constrained version, a few
more arbitrary input parameters over and above to that of the SM. Thus, it
has become imperative to constrain the SUSY parameter space as far as possible
from existing data. 

There are two aspects to this practice. First, take the experimental data and
find how much space we can allow for the SUSY parameters. Second, take the 
bounds obtained by method 1 and see what signals one should observe in
present or future experiments (and try to explain if there are any apparent
anomalies in the present data). Taken together, these two methods form a 
strong tool to observe indirect SUSY (or for that matter, any NP) signals,
which is complementary to the direct observation of the new particles in
high-energy machines like the Large Hadron Collider (LHC).

Since SUSY breaks at a high scale, there is no compelling mechanism to
suppress the flavor-changing neutral current (FCNC)
effects once we take the renormalization group (RG) 
evolutions into account. The problem is most severe for the gravity-mediated
SUSY breaking (SUGRA) type models. A huge number of models have been proposed
to solve this problem; we will not go into them here. Rather, we will 
focus on an equally important area of study, {\em viz.}, constraining the
FCNC parameters from experimental data. This shows which models are to stay
and which are not. 

The problem is worse in the RPV version of SUSY, since there exists a large
number of FCNC couplings from the very beginning. There are no theoretical
limits on these couplings except that they better not be nonperturbative
even at the scale of the Grand Unified Theories (GUT). The only way
to constrain the individual couplings and the products of two (or more)
of them is from experimental data. 

We will just discuss, as a sample process, the mixing of neutral B mesons,
and the effects of RPC and RPV SUSY on it (more processes will
be discussed in a subsequent publication). Why this process? The reasons 
are manyfold: (i) The theoretical part of SM and SUSY are both well-known
(including higher-order QCD corrections), 
apart from the uncertainties in some of the inputs; 
(ii) The mixing and the CP-asymmetry data have reached sufficient precision, 
and order-of-magnitude improvements are likely to occur in near future; (iii)
The SM amplitude is one-loop, so that the RPC SUSY amplitude, which must be 
a one-loop process, has a fair chance of competing, even with high sparticle
masses; (iv) The RPV SUSY amplitude is also one-loop and competes on the
same ground (there may be even tree-level amplitudes, but the couplings are
highly suppressed); (v) There are some decay channels, {\em e.g.},
$B\r \phi K$ \cite{phiks}, $B\r \eta' K$ \cite{etaprimek}, $B\r\pi\pi$
\cite{pipi}, which indicate that there
may be signals of NP hidden in B decays (although a SM explanation is never
ruled out).

Effects of RPC SUSY in $\bbbar$ mixing have been exhaustively studied in
the literature, and constraints were put on the FCNC
parameters of different SUSY models \cite{gerard,gabbiani,becirevic}. 
Apart from that, SUSY effects on various B decay processes (radiative,
leptonic, semileptonic and nonleptonic)
have been thoroughly investigated, but we are not going to discuss this
aspect in the present paper \cite{kundu-pramana}. 
A similar exercise has been performed for RPV SUSY models too \cite{rpv:b,
gg-arc}. 

In this paper, we will explore the robustness of some of the FCNC
parameters in RPC and RPV SUSY as quoted in the
literature \cite{gabbiani,becirevic,gg-arc}. 
The importance of this is threefold: first, this will serve as
an update of the existing results in the light of new data; second, this will
show that the bounds can get substantially relaxed once
we take all the SM uncertainties into account (a systematic study of the SUSY
FCNC parameters, taking this point consistently into account, has not been
performed as far as we know, though Ref.\ \cite{becirevic} does it in a 
sufficiently exhaustive way for RPC
SUSY only); and third, since the constraints
obtained in this paper are the most conservative ones, it should tell the
experimenters what sort of signal are to be expected at the most. 
We will also show how robust these bounds are if there happens to be a 
{\em natural} cancellation between RPC and RPV SUSY effects.

Here one must note the limitations of the B-related experiments in
constraining the NP models. Apart from the experimental uncertainties
($\sin(2\beta)$, CKM elements, branching fractions, etc.), there are
a number of inherent theoretical uncertainties, most of which stem from
the nonperturbative nature of low-energy QCD. To disentangle signatures
of NP, one must be fairly lucky to get a sizable deviation from the
SM prediction. This is precisely the reason why signatures of RPC SUSY
will be unobservable in those decay modes which have a tree-level
amplitude in the SM.

We will consider only the gluino-mediated box for the RPC SUSY amplitude.
This constrains the mixing parameters $\delta^d_{13}$ 
between the first and the third 
generations for the down-quark sector. Though neutralino and charged-Higgs
boson mediated boxes are expected to be small compared to the gluino box, an
almost equally large contribution comes from the chargino diagram 
\cite{khalil,debrupa}. Since the latter diagram constrains the mixing in the 
up-quark sector (the $\delta^u_{13}$ parameters), 
it will not be relevant for our future discussion.
The relevant details are to be found in Section 2. For RPV SUSY we will
consider only one product coupling giving rise to a new mixing amplitude
to be nonzero at a time.

We will assume, just for simplicity, that NP affects only the mixing 
amplitude, but there is no NP in the subsequent decay processes. On the one
hand this means that the CP asymmetry in the channel $B\r J/\psi K_S$
measures the phase in the box amplitude. This is easy to implement for
RPV SUSY by choosing appropriate combinations of nonzero flavor-dependent
couplings \footnote{However, this is true only for channels like $B\r
J/\psi K_S$. The channel $B\r\pi^+\pi^-$, which measures $\alpha$, has
RPV contributions both in mixing and in subsequent decay, and they must be
treated together \cite{bdk2}.}. For RPC SUSY, this assumption amounts to
neglecting loop-induced SUSY contributions to tree-level SM ones, which is
a safe assumption.

The QCD corrections to the box amplitude has been implemented upto 
Next to Leading Order (NLO), both for RPC and RPV SUSY. The relevant 
anomalous dimension matrix may be found in  \cite{becirevic}.

The paper is organized as follows: in Section II, we briefly recall the FCNC
phenomena in RPC and RPV SUSY, and discuss our numerical inputs in Section III. 
 Section IV deals with the constraints coming from
RPC SUSY, while Section V does the same job for the RPV version. In Section
VI we summarize and conclude.

\section{\boldmath$\bbbar~$\unboldmath Mixing in SUSY}
\subsection{\boldmath$\bbbar~$\unboldmath  in the SM}

The off-diagonal element in the $2\times 2$ effective Hamiltonian for the
neutral B system causes the mixing between the gauge eigenstates $B^0
(\equiv \bar{b}d)$ and $\bar{B^0} (\equiv b\bar{d})$ \cite{buras-fleischer}. 
The mass difference
between the two mass eigenstates $\Delta M_d$ is given by
\be
\Delta M_d = 2|M_{12}|
\ee
where
\bea
M_{12}&\equiv& {\bra \bar{B^0}|H_{eff}|B^0\ket\over  2m_B}\nonumber\\
&=& {G_F^2\over 6\pi^2}(V_{td}V_{tb}^*)^2
\eta_B m_B f_B^2 B_B m_W^2 S_0(x_t) ,
\eea
with $x_t = m_t^2/m_W^2$, and 
\be
S_0(x) = {4x-11x^2+x^3\over 4(1-x)^2} - {3x^3\ln x \over 2(1-x)^3}.
\ee
We follow the convention of Ref.\ \cite{buras-fleischer} for normalization
of the meson wave functions.
The perturbative QCD corrections are parametrized by $\eta_B$, while the
nonperturbative corrections are dumped in $B_B$. $f_B$ is the B meson decay
constant. The subleading boxes with two charm quarks or one charm and one top
quark are entirely negligible (the same holds for RPC SUSY;
however, due to the nonuniversal nature of
the relevant couplings, this is not true for the RPV version).

For B decays to a flavor-blind final state $f$ (e.g., $\psiks$)
where there are no nonzero CKM phases in the decay amplitude, the
measured CP asymmetry is proportional to the imaginary part of the
mixing amplitude. For $\bbbar$ box, this is $\sin(2\beta)$ as $arg(V_{td})
= -\beta$ (we will implicitly assume the Wolfenstein parametrization of
the CKM matrix, though the physical observables are parametrization
invariant). For $B_s$ box, the amplitude is real, so there is no
CP violation in the SM (to the leading order, {\em i.e.}, neglecting ${\cal
O}(\l^4)$ terms in the CKM matrix, where $\l = V_{us} \approx 0.22$).

NP adds up one (or more) new term to $M_{12}$. Even if it is real, 
the effective phase $\beta_{eff}$ should change from $\beta$.  
Thus, $\Delta M_d$ and $A_{CP}(B\r\psiks)$, taken together, should 
constrain both the real and the imaginary parts of the NP amplitude. 

Suppose there is one NP amplitude with a weak phase of $-2\phi$ so that one
can write
\be
M_{12} = |M_{12}^{SM}|\exp(-2i\beta) + |M_{12}^{NP}|\exp(-2i\phi).
\ee
This immediately gives the effective mixing phase $\beta_{eff}$ as
\be
\beta_{eff} = 0.5\arctan {|M_{12}^{SM}|\sin(2\beta) + |M_{12}^{NP}|\sin(2\phi)
\over |M_{12}^{SM}|\cos(2\beta) + |M_{12}^{NP}|\cos(2\phi)}
\ee
and the mass difference between the $B$ meson mass eigenstates as
\be
\Delta M_d = 2\left[|M_{12}^{SM}|^2 + |M_{12}^{NP}|^2 + 2|M_{12}^{SM}|
|M_{12}^{NP}|\cos 2(\beta-\phi)\right]^{1/2}.
\ee
These are going to be our basic formulae. The only remaining task is to
find $M_{12}^{NP}$. 

Note that even if $\phi = 0$, $\beta_{eff}\not=
\beta$, which means that even for real NP amplitudes, the constraints
vary with the choice of $\beta$. The CP asymmetry in $B\r \psiks$ measures
$\beta_{eff}$. However, with NP contributing to $\Delta M_d$, the $\bbbar$
mixing input to the standard CKM fit is lost, and $\beta$ essentially
becomes a free parameter. The same happens for $V_{td}$ also. We will 
discuss these issues in Section III.

\subsection{R-Parity conserving SUSY}

\begin{figure} 
\begin{center}  
\centerline{\hspace*{3em}
\epsfxsize=14cm\epsfysize=4.5cm
                     \epsfbox{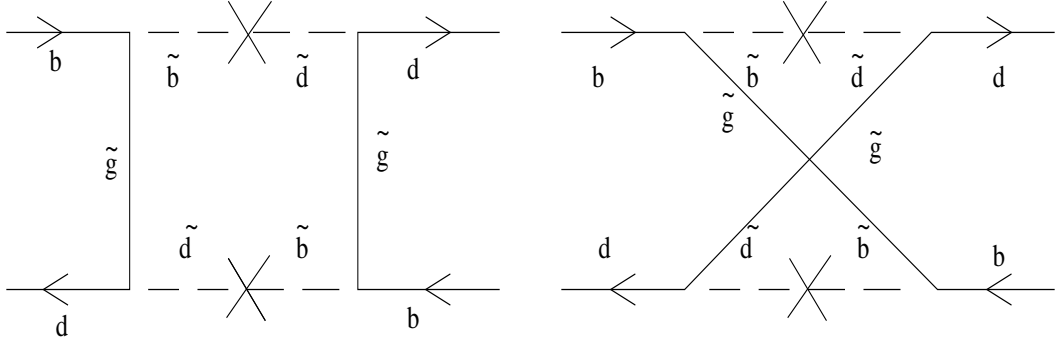}
}
\end{center}
\hspace*{-3cm}
  \caption{\em Gluino-mediated SUSY contributions to $\bbbar$ mixing.
One needs to add the crossed diagrams too. The crosses on the squark
propagators denote the insertion of the relevant $\delta$ parameters.
All chirality combinations are possible for the quarks, which we do not
show explicitly.}
\end{figure}

In RPC SUSY models there can be two more independent phases $\phi_A$ and
$\phi_B$ apart from the CKM phase, but the electric dipole moment of the
neutron constrains them to be small ($\sim{\cal O}(10^{-2}$-$10^{-3}$)
unless the squarks are extremely heavy or there is a fine-tuning 
betwen them \cite{ellis}. 
We take both of them to be zero, a choice which can be theoretically
motivated, since $\phi_A(\phi_B)$ is the relative phase between the 
common trilinear $A$-term (bilinear B-term) and the common gaugino mass
$M$ \cite{phasezero}. Even then one can have new contribution to CP
violation coming from SUSY FCNC effects \cite{gerard}.
The origin of SUSY FCNC can be easily understood: quark and squark mass 
matrices are not simultaneously diagonalizable. At $q^2\sim m_W^2$, 
radiative corrections induced by up-type (s)quark loops are important.
These corrections are typically of the order of log$(\Lambda_S/m_W)$
($\Lambda_S$ is the SUSY breaking scale) and hence can be large for SUGRA 
type models. This generates FCNC which occurs even in the
quark-squark-neutral gaugino vertices, but the flavor structure is controlled
by the CKM matrix. However, this last feature
need not be true in any arbitrary SUSY
model, particularly those with nonuniversal mass terms.

One generally works in the basis where the quark fields are eigenstates of the
Hamiltonian. SUSY FCNC can be incorporated in two ways: (i) Vertex mixing,
an approach where the squark propagators are flavor and `chirality' 
\footnote{Squarks cannot have a chirality, but this is just a loose term to
denote the partners of respective chiral fermions.} conserving, and the
vertices violate them; (ii) Propagator mixing, where flavor and `chirality'
are conserved in the vertices but changed in propagators. The second approach
is more preferred for phenomenological analysis, since the higher order QCD
corrections are known better. This is also known as the Mass Insertion
Approximation (MIA) \cite{hall,gabbiani}.

At the weak scale one can write the $6\times 6$ squark mass matrix (say the
down type) as
\be
{\tilde{\cal M}_D}^2 = \pmatrix {{\tilde{\cal M}_{DLL}}^2|^{tree} 
+ \Delta_{LL}^2 & \Delta_{LR}^2\cr
 \Delta_{RL}^2 & {\tilde{\cal M}_{DRR}}^2|^{tree} + \Delta_{RR}^2}
\ee
where the $\Delta$ terms incorporate the FCNC effects. Different FCNC
effects are parametrized in terms of $\delta^{ij}_{AB}\equiv \Delta^{ij}_{AB} 
/{\tilde m}^2$, where $\tilde m = \sqrt{m_1m_2}$, the geometric mean of
the masses of the two participating squarks, $i$ and $j$ are flavor
indices, and A and B are chiral indices. The $\delta$s are completely
calculable in constrained MSSM, but this is not true in general. Theoretically,
for the success of perturbative analysis, one expects $|\delta|<1$. The 
gluino-mediated box diagrams causing $\bbbar$ mixing constrain only
$\delta^{13}$s of different chiralities in the down-quark sector. (Thus, 
$\delta^{13}_{AB}$ in our notation means $(\delta^d_{13})_{AB}$ in
the more conventional notation; there is, however, no chance of confusion
since we do not deal with the $\delta^u$ parameters.) The same thing can
be said for the up-quark sector by considering the chargino
contributions \cite{khalil}.

The $\Delta B = 2$ effective Hamiltonian can be written in a general form as
\be
{\cal H}_{eff}^{\Delta B = 2} = \sum_{i=1}^5 c_i O_i + \sum_{i=1}^3
\tilde c_i \tilde O_i + H.c.
\ee
where
\bea
O_1&=&(\bar b \gamma^\mu P_L d)_1(\bar b \gamma_\mu P_L d)_1,\nonumber\\
O_2&=&(\bar b P_R d)_1(\bar b  P_R d)_1,\nonumber\\
O_3&=&(\bar b P_R d)_8(\bar b  P_R d)_8,\nonumber\\
O_4&=&(\bar b P_L d)_1(\bar b  P_R d)_1,\nonumber\\
O_5&=&(\bar b P_L d)_8(\bar b  P_R d)_8,
   \label{operators}
\eea
and $P_{R(L)}=(1+(-)\gamma_5)/2$. The subscripts 1 and 8 indicate whether
the currents are in color-singlet or in color-octet combination. The $\tilde
O_i$s are obtained from corresponding $O_i$s by replacing $L\leftrightarrow R$.

The Wilson coefficients have been computed
at the high scale $M_S$ (chosen to be the arithmetic mean of the average
squark mass and the gluino mass)
by evaluating the diagrams in Fig.\ 1. We quote
the results \cite{gabbiani}:
\bea
c_1&=&-R\left(24 x f_6(x) + 66 \tilde f_6(x)\right)(\delta^{13}_{LL})^2,
\nonumber\\
c_2&=&-R\left(204 x f_6(x)\right)(\delta^{13}_{RL})^2,
\nonumber\\
c_3&=&R\left(36 x f_6(x)\right)(\delta^{13}_{RL})^2,
\nonumber\\
c_4&=&-R\left\{
\left(504 x f_6(x) - 72 \tilde f_6(x)\right)\delta^{13}_{LL}\delta^{13}_{RR}
-132 \tilde f_6(x) \delta^{13}_{LR}\delta^{13}_{RL}\right\},
\nonumber\\
c_5&=&-R\left\{
\left(24 x f_6(x) +120 \tilde f_6(x)\right)\delta^{13}_{LL}\delta^{13}_{RR}
-180 \tilde f_6(x) \delta^{13}_{LR}\delta^{13}_{RL}\right\}.
\eea
Here $R = \alpha_s^2/216 m_{\tilde q}^2$ and $x = m_{\tilde g}^2 / 
m_{\tilde q}^2$.
The coefficients $\tilde c_i$s can be obtained from corresponding $c_i$s
again with $L\leftrightarrow R$. The functions $f_6$ and $\tilde f_6$ are
given by
\bea
f_6(x)&=&{6(1+3x)\ln x+x^3-9x^2-9x+17\over 6(x-1)^5},\nonumber\\
\tilde f_6(x)&=&{6x(1+x)\ln x-x^3-9x^2+9x+1\over 3(x-1)^5}.
\eea
Next, one should evolve these coefficients down to the low-energy scale,
taken, following Ref.\ \cite{becirevic}, to be $\mu = m_b = 4.6$ GeV, using
the NLO-QCD corrections. The low-scale Wilson coefficients are
\be
c_i(\mu) = \sum_r \sum_s \left(b_r^{i,s} + \eta c_r^{i,s}\right)\eta^{a_r}
c_s(M_S)
\ee
where $\eta = \alpha_s(M_S)/\alpha_s(m_t)$. For the numerical values of
$a$, $b$ and $c$ matrices we refer the reader to eq.\ (10) of Ref.\
\cite{becirevic}.

The operators $O_i$ are also to be renormalized at the scale $\mu$. 
The expectation values of these operators between $\bar{B^0}$ and $B^0$ 
at the scale $\mu$ are given by
\bea
\bra O_1(\mu)\ket &=& {2\over 3} m_B^2 f_B^2 B_1(\mu),\nonumber\\
\bra O_2(\mu)\ket &=& -{5\over 12} S m_B^2 f_B^2 B_2(\mu),\nonumber\\
\bra O_3(\mu)\ket &=& {1\over 12} S m_B^2 f_B^2 B_3(\mu),\nonumber\\
\bra O_4(\mu)\ket &=& {1\over 2} S m_B^2 f_B^2 B_4(\mu),\nonumber\\
\bra O_5(\mu)\ket &=& {1\over 6} S m_B^2 f_B^2 B_5(\mu),
\eea
where
\be
S = \left({m_B\over m_b(m_b) + m_d(m_b)}\right)^2
\ee
The $B$-parameters, whose numerical values are given in Section 3, have been
taken from \cite{lattice-b}. Note that the expectation values are scaled by
factor of $2m_B$ over those given in some literature due to our different
normalization of the meson wavefunctions. It is trivial to check that 
both conventions yield the same values for physical observables. 

We wish to draw the reader's attention to the fact that with changing $x$,
the interference pattern between the SM box and the SUSY box changes.
For example, if only $(\delta^{13}_{LL})^2$ is nonzero, there is a 
constructive interference with the SM box if $x<2.1$, and a destructive 
interference otherwise. This is just because $c_1$ changes sign as we go to
higher values of $x$. Thus, if one chooses $(\delta^{13}_{LL})^2$ to be real,
$\beta_{eff}$ goes down from $\beta$ for low $x$, and goes up for
high $x$. Near the crossover region, the SUSY contribution almost vanishes,
so one can in principle have large $\delta$ parameters. We do not analyze
these regions since they smell of a fine-tuning, but one should keep this
point in mind.

\subsection{R-Parity violating SUSY}

\begin{figure}
\begin{center}
\centerline{\hspace*{3em}
\epsfxsize=14cm\epsfysize=4.5cm
                     \epsfbox{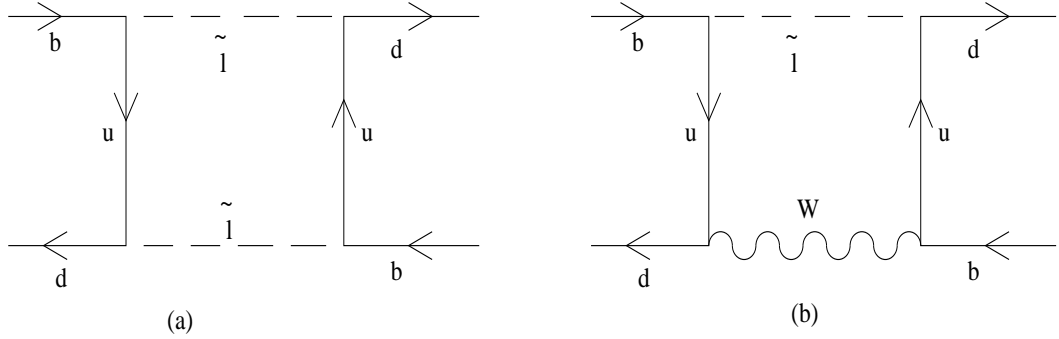}
}
\end{center}
\hspace*{-3cm}
  \caption{\em R-parity violating contributions to $\bbbar$ mixing. 
Figure (a) corresponds to L4, while figure (b) to L2 amplitudes (see text for
their meanings).
One must add the crossed diagrams, as well as the diagrams in (b) where
the $W$ is replaced by the charged Higgs or the charged Goldstone. The internal
slepton can be of any generation, and so can be the internal charge $+2/3$
quarks, generically depicted as $u$.}
\end{figure}

R-parity is a global quantum number, defined as $(-1)^{3B+L+2S}$, which
is $+1$ for all particles and $-1$ for all superparticles. In the minimal
version of supersymmetry and some of its variants, R-parity is assumed
to be conserved {\em ad hoc}, which prevents single creation or
annihilation of superparticles. However, models with broken R-parity can
be constructed naturally, and such models have a number of interesting
phenomenological consequences \cite{rpvrefs1,rpvrefs2}. 
Some of these R-parity violating models can be motivated from an underlying
GUT framework \cite{rpvgut}.

It is well known that in order to avoid rapid proton decay one cannot
have both  lepton number and  baryon number violating RPV model, and we 
shall work with a lepton number violating one. This leads
to slepton/sneutrino mediated B decays, and new amplitudes for $\bbbar$
mixing with charged sleptons and up-type quarks (and maybe $W$, charged Higgs
and charged Goldstone bosons) flowing inside the loop (see Fig.\ 2). 
Since the current lower bound on the
slepton mass is weaker than that on squark mass,
larger effects within the reach of current round of experiments are
more probable in this scenario. We start with the superpotential
\be
\label{w}
{\cal W}_{\lambda'} = \lambda'_{ijk} L_i Q_j D^c_k,
\ee
where $i, j, k = 1, 2, 3$ are quark and lepton generation indices;
$L$ and $Q$ are the $SU(2)$-doublet lepton and quark superfields and
$D^c$ is the $SU(2)$-singlet down-type quark
superfield respectively. Written in terms of component fields, this 
superpotential generates six terms, plus their hermitian conjugates,
but for our present purpose the only relevant term is
\be
{\cal L}_\rnot\supset -\lambda'_{ijk}\tilde e^i_L \bar{d}^k_Ru^j_L + H.c.
\ee 
With such a term, one can have two different kind of boxes, shown in Fig.\ 2,
that contribute
to $\bbbar$ mixing: first, the one where one has two sleptons flowing inside
the loop, along with two up-type quarks \cite{decarlos-white}, and secondly, 
the one where one slepton, one $W$ (or charged Higgs boson or Goldstone 
boson) and two
up-type quarks complete the loop \cite{gg-arc}. It is obvious that the first
amplitude is proportional to the product of four $\lambda'$ type
couplings, and the second to the product of two $\lambda'$ type couplings
times $G_F$. We call them $L4$ and $L2$ boxes, respectively, for brevity.

The effective Hamiltonian for the $L4$ boxes reads (no sum over $k$)
\bea
{\cal H}_{L4} = {({\l'_{ik1}}^*\l'_{ik3})^2\over 32 \pi^2
{\tilde m_l}^2} I\left({m_{q_k}^2\over {\tilde m_l}^2}\right) \tilde{O_1}
\eea
where $\tilde m_l$ is the slepton mass, and $\tilde O_1$ has been defined in
eq.\ (\ref{operators}). 
Actually, there can be different up-type quarks and different 
generations of sleptons flowing in the box, but that makes the Hamiltonian
proportional to the product of four $\l'$ couplings, which we avoid
for simplicity. The function $I(x)$ is given by
\be
I(x) = {1-x^2+2x\log x\over (1-x)^3}.
\ee
For $L2$ boxes, there are three different types of amplitudes in the 
Feynman gauge: involving, along with a slepton, a $W$, a charged Higgs
boson, or a charged Goldstone boson. The sum is given by \cite{gg-arc}
\be
{\cal H}_{L2} = {G_F {\l'}_{ik1}^*{\l'}_{ik3}\over 4\sqrt{2}\pi^2} V_{u_kb}^*
V_{u_kd} \left[ (1+\cot^2\beta) x_k^2 J(x_k) + I(x_k)\right] O_4
\ee
where $x_k = m_{u_k}^2/m_{\tilde {m_l}}^2$,  
\be
J(x) = {-2(x-1)+(x+1)\log x\over (x-1)^3}
\ee
and $\cot\beta = v_d/v_u$, the ratio of the vacuum expectation values of
the two Higgses that give mass to the down- and the up-type quarks
respectively (not to be confused with the phase of $V_{td}$). 
The Hamiltonian is slightly modified if one has different up-type
quarks $u_k$ and $u_l$ in the box:
\be
{\cal H}_{L2} = {G_F {\l'}_{ik1}^*{\l'}_{ip3}\over 4\sqrt{2}\pi^2} V_{u_kb}^* 
V_{u_pd} \left[ (1+\cot^2\beta) x_k x_p K(x_k) + L(x_k)\right] O_4
\ee
where we have assumed $m_k > m_p$; if not, the arguments of $K$ and $L$
are to be replaced by $x_p$. The functions are given by
\bea
K(x) &=& {x-1-\log x\over (x-1)^2},\nonumber\\
L(x) &=& 1 - xK(x).
\eea
Note that one can have an imaginary part in the amplitude when the internal 
quarks are both light, but we neglect that effect for our present 
purpose \footnote{Even from dimensional arguments, the ratio of the 
imaginary and the real parts of the mixing amplitude should at most be of 
the order of $m_b^2/m_t^2$ \cite{bss}, 
and hence the phase introduced by such an imaginary part
can be neglected in the analysis.}.
Also note that in the light of LEP data which definitely favours $\tan\beta
\geq 2-3$ \cite{lepsusy}, the Goldstone contributions are dominant over the
charged Higgs contributions, which are suppressed by $\cot^2\beta$. In deriving
the above expressions, we have assumed all scalars flowing inside the box to 
have equal mass.

In general both the $\l'$ type couplings can have a phase, but one of them
can be absorbed in the definition of the slepton propagator, so it is
enough to consider the one remaining phase.

It is easy to see that the relevant equations  (4)-(6)
need to be slightly modified to include two
NP amplitudes when same quarks flow in the loop; this being a trivial
exercise, we do not show the formulae explicitly. For small values of the
corresponding $\l'$ couplings, $H_{L2}$ dominates $H_{L4}$, but the role
may get reversed for large values of the product coupling. Thus, one gets
two bands in the RPV coupling versus $\Delta M_d$ plane, and our bounds
correspond to the outer band. There is no such complication when only the 
$L2$ amplitude is present. 
 
\section{Numerical Inputs}

%
\begin{table}[htbp]
\begin{center}
\begin{tabular}{||c|c|c||}
\hline
 && \\
Quantity & Value & Remarks\\
\hline
$m_B$        & 5.2794 GeV &  \cite{pdg2002}\\
$m_W$        & 80.423 GeV &  \cite{pdg2002}\\ 
$\Delta M_d$ & $0.502\pm 0.007$ ps$^{-1}$   & \cite{heavyflav} \\
$\sin(2\beta_{eff})$     & $0.681$-$0.784$ & \cite{ckmfitter2002}, at $1\sigma$
C.L. \\
$\beta_{eff}$ & $21.4^\circ$-$26.0^\circ$ &  Assumed to be $< 90^\circ$      \\
$\gamma$ & $44^\circ$-$72^\circ$ & \cite{ckmfitter}, at $1\sigma$ C.L.  \\
$m_t^{\bar{MS}}(m_t^{\bar{MS}})$ & 166 GeV         & \cite{ckmfitter}  \\
$m_b^{\bar{MS}}(m_b^{\bar{MS}})$ & 4.23 GeV       & \cite{becirevic}  \\
$m_b(m_b)$                       &  4.6 GeV &   \\
$m_d(m_b)$                       &  5.4 MeV &   \\
$\eta_B$   & $0.55\pm 0.01$ & \cite{ckmfitter} \\
$f_B\sqrt{B_B}$ & $230 \pm 28  \pm 28$ MeV & \cite{ckmfitter}\\
$\alpha_s(m_Z)$     & $0.1172\pm 0.002$   &  \cite{pdg2002}\\
$|V_{td}|\times 10^3$ & 6.3 - 9.6 & \cite{ckmfitter-old}, at 95\% C.L. \\
$|V_{ub}|\times 10^3$ & 2.49 - 4.55 & \cite{ckmfitter-old}, at 95\% C.L. \\
\hline
\end{tabular}
\caption{Input parameters used for the numerical analysis. }
\end{center}
\end{table}

The important numerical inputs used in our work is shown in Table 1. A few
points are to be noted.

Unless shown in the table, we have not taken the experimental uncertainty of a 
quantity into account. For example, we have used the central values of
the CKM elements, except that of $V_{td}$ and $V_{ub}$, in our analysis.
They are taken from Ref.\ \cite{ckmfitter-old}, extracted without the
world average of $\sin(2\beta)$. This is justified since now $\sin(2\beta)$
itself has a NP contribution, and so should not be used to extract 
the values of the CKM parameters.

$V_{td}$ is determined from $\Delta M_d$; thus, the SM fit for $V_{td}$
no longer works when there is one or more NP amplitudes to the box. Since
the CP asymmetries are controlled by phases not all of which are in the CKM
matrix, the usual argument of the so-called Universal Unitarity Triangle
(UUT) \cite{uut} 
does not hold. In essence $V_{td}$ becomes a free parameter, only
controlled by the unitarity of the CKM matrix. To take into account this
feature, we have taken the 95\% confidence limit (CL) for $V_{td}$,
extracted without the global average of $\sin(2\beta)$, for our analysis
\cite{ckmfitter-old}. The same holds for $V_{ub}$, which contains the phase
$\gamma$. As pointed out by \cite{becirevic}, $\gamma$ also becomes a free
parameter. We address this issue by keeping the range of $\gamma$ within 
1$\sigma$ CL quoted by \cite{ckmfitter}, while $V_{ub}$ is varied over its
95\% CL range. This, we have checked, essentially covers the whole region
generated by a narrower range of $V_{ub}$ and a wider range of $\gamma$
covering $0$ to $2\pi$, with the constraint that the three-generation 
CKM matrix is unitary ({\em i.e.}, the unitarity triangle should close). 
The bounds are of the same order, but slightly 
more conservative for the former case.

The imaginary part of $\delta$s crucially depends on the choice of $\beta$.
For our analysis, we have varied $\sin(2\beta)$ between 0
and 1, putting a special emphasis on those values which make $\beta
\approx\beta_{eff}$. 

The most important parameter for the RPC SUSY analysis is the average squark
mass, which we fix at 500 GeV, effectively neglecting the splitting caused
by the SU(2) D-term.
The gluino mass is varied between
$0.3<m_{\tilde g}^2/m_{\tilde q}^2 < 4.0$. The bounds more or less scale
with the squark mass, as can be seen from eqs.\ (8) and (10). This, however,
leaves out some models, {\em e.g.}, those with an extremely light gluino.

For RPV SUSY analysis, we take all sleptons to be degenerate at 100 GeV.
The bounds on the product couplings scale as the square of the slepton
mass. The charged Higgs is also assumed to be at 100 GeV. As is discussed
below, the precise value of the Higgs mass is not a crucial input. 

The value of $\tan\beta$ (the SUSY parameter)
is fixed at 3, compatible with the lower bound of the
recent LEP analysis \cite{lepsusy}.  A glance at eqs.\ (19) and (21)
should convince the reader
that the bounds are not very sensitive to the exact value of $\tan\beta$,
unless it is small,
since the Goldstone contributions independent of $\beta$ control the 
show. We have explicitly checked the robustness of the bounds with the
variation of $\tan\beta$.

The $B$-parameters have been taken from \cite{lattice-b}, and at $\mu=m_b$,
read
\be
B_1 = 0.87(4)^{+5}_{-4},\ B_2 = 0.82(3)(4),\
B_3 = 1.02(6)(9),\ B_4 = 1.16(3)^{+5}_{-7},\
B_5 = 1.91(4)^{+22}_{-7}.
\ee
The low-energy Wilson coefficients can be found in \cite{becirevic}.
 
\section{Results for RPC SUSY}

%
\begin{table}[htbp]
\begin{center}
\begin{tabular}{||c|c|c|c|c|c||}
\hline
$\sin 2\beta$ & x & 
$\sqrt{|{(\delta^{13} _{LL})^2}|}$  & 
 $\sqrt{|(\delta^{13} _{LR})^2|}$ &
$\sqrt{|\delta^{13} _{LL}\delta^{13} _{RR}|}$ &
 $\sqrt{|\delta^{13} _{LR}\delta^{13} _{RL}|}$ \\
\hline
    & 0.3 & 0.046 & 0.021 & --- & --- \\
1.0 & 1.0 & 0.099 & 0.023 & --- & --- \\
    & 2.0 & 0.27 & 0.026 & --- & --- \\
    & 4.0 &  ---  & 0.031 & --- & --- \\
\hline
      & 0.3 & 0.017 & 0.0075 & 0.0030 & 0.0040\\
0.732 & 1.0 & 0.036 & 0.0080 & 0.0033 & 0.0068\\
      & 2.0 & 0.10 & 0.0095 & 0.0039 & 0.0099\\
      & 4.0 & 0.090 & 0.012 & 0.0048 & 0.015\\
\hline
      & 0.3 &  ---  &  ---   & 0.0078 & 0.011\\
0.5   & 1.0 &  ---  &  ---   & 0.0088 & 0.018\\
      & 2.0 &  ---  &  ---   & 0.010 & 0.026\\
      & 4.0 & 0.235 &  ---   & 0.012 & 0.039\\
\hline
\end{tabular}
\caption{Bounds on the $\delta^{13}$ parameters when they are all real.}
\end{center}
\end{table}

The real and imaginary parts of the SUSY amplitude being established through
the real and imaginary parts of the corresponding $\delta$s, it is easy to
find the limits on those real and imaginary parts. We perform a scan on
the complete range of $V_{td}$ and $f_B \sqrt{B_B}$, as well as on the
SUSY phase $\phi$, where generically $\delta^2 = |\delta^2|\exp(-2i\phi)$,
over the range 0 to $\pi$. We demand that $\Delta M_d$ and $\sin(2\beta_{eff})$
should lie between the values specified in Table 1. The results are obtained
for various values of $\beta$, even for extreme values like $\sin(2\beta)=0$
or 1.

%
\begin{table}[htbp]
\begin{center}
\begin{tabular}{||c|c|c|c|c|c||}
\hline
$\sin 2\beta$ & x & 
$\sqrt{|Re {(\delta^{13} _{LL})^2}|}$  & 
$\sqrt{|Im {(\delta^{13} _{LL})^2}|}$  & 
$\sqrt{|Re {(\delta^{13} _{LR})^2}|}$  & 
$\sqrt{|Im {(\delta^{13} _{LR})^2}|}$  \\ 
\hline
    & 0.3 & 0.046 & 0.079  & 0.021 & 0.035 \\
1.0 & 1.0 & 0.10  & 0.18   & 0.022 & 0.039 \\
    & 2.0 & 0.27  & 0.27   & 0.026 & 0.046 \\
    & 4.0 & 0.23  & 0.30   & 0.031 & 0.061 \\
\hline
      & 0.3 & 0.077 & 0.078  & 0.028  & 0.030 \\
0.732 & 1.0 & 0.16  & 0.16   & 0.035  & 0.038 \\
      & 2.0 & 0.27  & 0.28   & 0.038  & 0.040 \\
      & 4.0 & 0.24  & 0.25   & 0.051  & 0.054 \\
\hline
      & 0.3 & 0.081 & 0.070  & 0.031  & 0.027 \\
0.5   & 1.0 & 0.17  & 0.15   & 0.038  & 0.032 \\
      & 2.0 & 0.31  & 0.23   & 0.046  & 0.038 \\
      & 4.0 & 0.28  & 0.19   & 0.053  & 0.044 \\
\hline
      & 0.3 & 0.084 & 0.048  & 0.034  & 0.021 \\
0.0   & 1.0 & 0.18  & 0.10   & 0.040  & 0.026 \\
      & 2.0 & 0.27  & 0.28   & 0.054  & 0.035 \\
      & 4.0 & 0.30  & 0.24   & 0.060  & 0.032 \\
\hline
\end{tabular}
\caption{Bounds on the real and the imaginary parts of $\delta^{13}_{LL}$ 
and $\delta^{13}_{LR}$.}
\end{center}
\end{table}

Tables 2 and 3 summarize our results. In table 2, we show the bounds
on various $\delta$ parameters (and their combinations) when they are
real, assuming only one to be nonzero at a time.
Note how the interference pattern changes for $\delta^{13}_{LL}$ with $x$;
it is constructive for small $x$, and destructive for large $x$. Near $x=2$,
the crossover occurs, so one can have a large value for $\delta^{13}_{LL}$.
A look at the respective Wilson coefficients and hence the interference
pattern should help the reader to understand the missing entries. Without
the CP-asymmetry constraint, nontrivial entries would occur everywhere
\cite{gabbiani}.

Table 3 shows the real and imaginary parts for $\delta^{13}_{LL}$ and
$\delta^{13}_{LR}$ for several values of $\sin(2\beta)$. Due to the presence
of the nonzero SUSY phase, all entries are nonvanishing. Also, the rise
of $\delta_{LL}^{13}$ near $x=2$ is less pronounced. We do not show
the entries for negative values of $\sin(2\beta)$; they are more or less equal
with their counterparts for positive $\sin(2\beta)$. There are a few
exceptions, which do not change the general result.
 
We caution the reader that these bounds are the most conservative ones at the
particular benchmark points that we have chosen, but by no means signify
an impossibility of having larger $\delta$s at other points in the SUSY
parameter space. Particularly, note that these bounds scale with the squark
mass.

\section{Results for RPV SUSY}

%
\begin{table}[htbp]
\begin{center}
\begin{tabular}{||c|c|c|c||}
\hline
$\lambda'\lambda'$ combination 
& $Re(\lambda'\lambda')$ & $Im(\lambda'\lambda')$ & Direct bound\\
\hline
(i31)(i33)& $1.8\times 10^{-3}$ &  $1.7\times 10^{-3}$ & 0.202 \\
(i21)(i23)&  $1.2\times 10^{-3}$ (0.022)&  $1.2 (1.4) \times 10^{-3}$ & 0.27\\
(i11)(i13)&  $2.4 (2.6)\times 10^{-3}$&  $2.5 (2.6)\times 10^{-3}$ & $
2.7\times 10^{-3}$ (*)\\
(i11)(i23)&0.016&0.016 & 0.057\\
(i11)(i33)& 0.026$^\dag$& 0.026$^\dag$ & 0.026\\
(i21)(i13)&$2.5 (3.0)\times 10^{-4}$&$2.8\times 10^{-4}$ & 0.057\\
(i21)(i33)&0.098&0.1 & 0.23\\
(i31)(i23)&$1.4\times 10^{-4}$&$ 1.35(1.6)\times 10^{-4}$ & 0.26\\

\hline

\end{tabular}
\caption
{Upper limits on the real and the imaginary parts of the relevant
$\lambda'\lambda'$ couplings for $\sin(2\beta) = 0.732$. 
The numbers in parenthesis show the maximum
possible value, for some other $\sin(2\beta)$. All the products, 
except the one marked with a dagger ($\dag$), show improvements over the
corresponding bounds obtained from direct product of the bounds of the
relevant $\lambda'$s (for $i=3$), shown in the fourth column. 
All the numbers in the fourth column are from \cite{allanach}, except the one
marked with asterisk, which is from \cite{bdk2}.
The bounds on the imaginary parts are new.} 
\end{center}
\end{table}

We explicitly assume that the contributions coming from the RPC SUSY
sector vanish if there is nonzero RPV interactions. This will be justified
{\em post hoc} when we discuss a sample case where both are present, and
there is a possibility of cancellation. 

The strategy is the same as that adopted for the previous subsection. 
The phase of the RPV product coupling is varied between $0$ and $2\pi$
while the magnitude of the coupling is assumed to be only positive. The
range of the scan is kept between the direct product limits,  {\em i.e.},
the limit one obtains when one multiplies the individual limits for the
two $\lambda'$ components. This limit, as can be easily checked 
\cite{allanach}, is most lenient for the third slepton generation. 
Only two of the relevant product couplings have been bounded from other
sources: the product $\l'_{i13}\l'_{i31}$ has a very stringent bound from
tree-level $\bbbar$ mixing, of the order of $10^{-8}$ (and hence we do not
discuss this product further), and the product $\l'_{i11}\l'_{i13}$ has been
constrained from the measured branching ratio and CP asymmetries in the 
$B\r\pi^+\pi^-$ channel \cite{bdk2} (marked with an asterisk in Table 4).
There is no bound on the imaginary parts of the couplings. 

\begin{figure}[htbp]
\vspace{-10pt}
\centerline{\hspace{-3.3mm}
\rotatebox{-90}{\epsfxsize=8cm\epsfbox{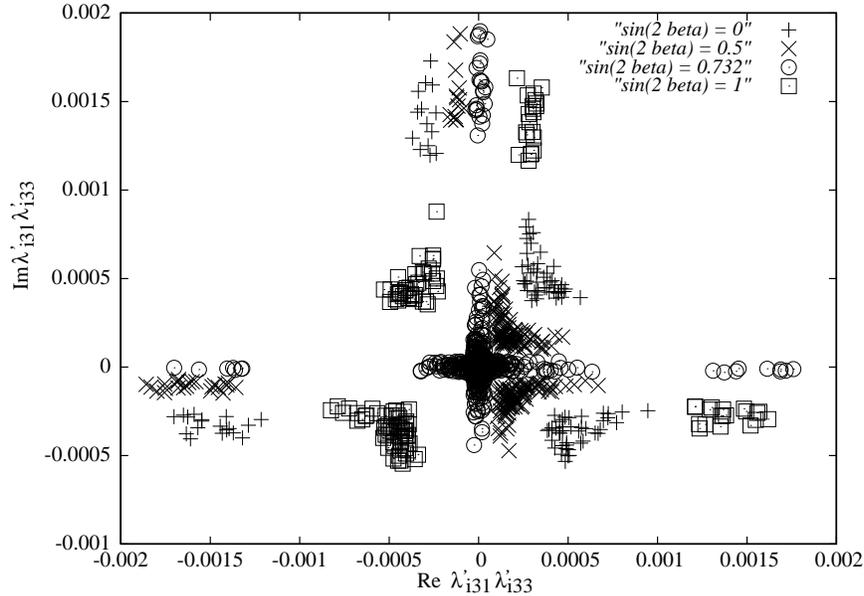}}}
\hspace{3.3cm}\caption[]{The real and imaginary parts of
$\l'_{i31}\l'_{i33}$ for various values of $\sin(2\beta)$.}
  \protect\label{fig3}
\end{figure}

Table 4 summarizes our results. Note that almost all the products, apart from
those two discussed above, and the one marked with a dagger,
have been improved, some by orders of magnitude. 
Six among these eight entries have been considered by the authors of 
\cite{gg-arc}. One may note that they obtained bounds which have the
same orders of magnitude to the ones that we get. However, there are
several ways in which we have improved upon their calculation, apart from
taking the updated data as input. These improvements are: (i) imposition of
the CP-asymmetry constraint, which was not available at their time (this
helps us to obtain the bounds on the imaginary parts of the couplings); (ii)
incorporation of the NLO QCD corrections, which are, anyway, expected to be
small --- at least they should not change the numbers by, say,
a factor of two; (iii) scan over the full range of the input parameters, as
we have already discussed; and (iv) consideration of the SM contribution,
and the possibilities of interference between SM, $L2$ and $L4$ amplitudes. 
Also note that they have computed the bounds for $\tan\beta = 1$.

Our bounds are more or less insensitive to the precise value of 
$\sin(2\beta)$ (the UT angle); 
the exceptions are shown in the table. Another feature is
that real and imaginary parts have almost the same bounds. Figure 3 
highlights this nature in detail for the product coupling $\l'_{i31}\l'_{i33}$.

Let us try to understand this figure. There are three main
regions, the first one (like a smeared cross) encompassing the origin, 
the second one, divided into four almost symmetrical fragments, 
around $|Re(\l'_{i31}\l'_{i33})| = 5\times 10^{-4}$, and a third fragmented
region about $|\l'_{i31}\l'_{i33}| = 15\times 10^{-4}$. It is the outermost
third region which gives the bound. It is easy to explain the origin of these
three regions, and sheds light on the role of $\sin(2\beta)$ in determining
the parameter space. 

The first one is governed by the SM, and the difference between the experiment
and the theory is filled up by RPV. Note that this region has points only for
$\sin(2\beta)=0.732$; thus, SM is allowed without any NP. There is a satellite
region, for $\sin(2\beta)=0.5$, where SM is not allowed (from the $A_{CP}$
constraint), and RPV fills in to generate the necessary CP asymmetry. 
The second region includes points for extreme values of $\sin(2\beta)$: 0 or 1.
SM is not allowed, and one needs a greater role from RPV to obtain the
observed CP asymmetry. Though both $L2$ and $L4$ boxes are allowed for this
RPV coupling, the $L2$ box dominates here. However, they come with opposite
sign, so there is a region where they interfere destructively (particularly
with the increase of the coupling), and RPV
contribution may even go to zero, leaving only SM. This generates the third
region and explains why one has points even for $\sin(2\beta)=0.732$.  

It is a possibility that both RPC and RPV SUSY are present, however 
pathological that may seem. Even in this extreme case the bounds are never
changed by orders of magnitude, unless there is a very precise fine-tuning.
For example, with only $\delta_{LL}^{13}$ and $\l'_{i21}\l'_{i33}$ present
(this particular combination is chosen since both of them have comparable
upper bounds), the bounds are relaxed to $\sqrt{|Re(\delta_{LL}^{13})^2|} =
0.22$, $\sqrt{|Im(\delta_{LL}^{13})^2|} = 0.26$, $Re(\l'_{i21}\l'_{i33}) =
0.12$, $Im(\l'_{i21}\l'_{i33}) = 0.11$, for $x=1$ and $\sin(2\beta) = 0.732$.
Thus we have reasons to be confident about these bounds. This is more so
if the limits have different orders of magnitude.

%
\def\r{\rightarrow}
\begin{table}[htbp]
\begin{center}
\begin{tabular}{||c|c|c|c||}
\hline
$\lambda'\lambda'$ combination 
& Decay channels & 
$\lambda'\lambda'$ combination 
& Decay channels\\
\hline
(i31)(i33)& $b\r d\bar{\ell_i} \ell_i$, $b\r d\bar{\nu_i} \nu_i$ &
(i21)(i23)& $b\r c\bar{c} d$, $b\r s\bar{s} d$\\
(i11)(i23)& $b\r c\bar{u} d$, $b\r s\bar{d} d$&
          & $b\r d\bar{\ell_i} \ell_i$, $b\r d\bar{\nu_i} \nu_i$ \\
(i21)(i13)& $b\r u\bar{c} d$, $b\r d\bar{s} d$&
(i11)(i13)& $b\r u\bar{u} d$, $b\r d\bar{d} d$\\
 & &
          & $b\r d\bar{\ell_i} \ell_i$, $b\r d\bar{\nu_i} \nu_i$ \\
\hline

\end{tabular}
\caption
{The possible decay channels of the B meson driven by the different
RPV product couplings. The final state mesons are not shown explicitly.
For the semileptonic decays, the outgoing leptons must be of the same
generation, denoted by $i$.
} 
\end{center}
\end{table}

Most of these product couplings contribute to various B-decay channels,
both nonleptonic and semileptonic. They are enlisted in Table 5. It is easy
to check which mesons and leptons come out in the final stage. The 
present $e^+e^-$ B factories, as well as the future hadronic machines, 
should put tighter constraints on these RPV product couplings. 

\section{Summary and Conclusions}

In this paper we have enlisted the constraints on the real and the imaginary
parts of the FCNC parameters of both R-parity conserving and R-parity
violating SUSY, coming from $\bbbar$ mixing. For RPC SUSY, these are the
conventional $\delta^d_{13}$ parameters of different chiralities. The same
analysis was performed by \cite{becirevic}; our results differ slightly from
theirs due to two reasons. We have performed a scan over all SM quantities,
including $V_{td}$ and $f_B^2B_B$, and our range of scan for $\gamma$ is
different from theirs. Weaker constraints on these $\delta$ parameters can
also be derived from the radiative decay $b\r d\gamma$.

For the RPV SUSY scenario, the FCNC parameters are the $\l'$ type lepton-number
violating couplings. One needs a product of two such couplings to generate
$\bbbar$ oscillation. There can be several such products, depending on
the choice of the quarks and sleptons flowing inside the box, whose real
and imaginary parts have been bounded from the experimental data. The bounds
on the real parts update the work of \cite{gg-arc} while the bounds on the
imaginary parts are derived for the first time in this paper. 

If we ever find a signal for NP in B factories, how can we be certain that 
it indeed comes from SUSY? There are three steps to ascertain that. First,
sort out those channels which show an abnormality. Second, Try to find the
model which can explain these anomalies. And third, check whether there are
other channels where one may expect to see an anomaly, and whether the anomaly
may be present in the data. If there is a prospective channel, one should
look for it. Confirmation of the nature of the NP is never possible without
the study of such correlated signals. Such correlated signals may even be
the direct production of new particles, {\em e.g.}, in the Large Hadron
Collider. A possible discriminating signal between RPC and RPV SUSY is
the fact that the RPV version may be highly flavor-specific, and so one
would expect the absence of anomaly in such channels which may be affected
in a more flavor-blind model such as the RPC SUSY.

For the help of our experimentalist colleagues, we enlist the possible
decay modes of the B meson which are driven by the RPV product couplings
discussed here. 
A careful study of the possible channels in present and future B factories 
should be able to put tighter constraints on the parameter space. Particularly,
the proposed higher luminosity $e^+e^-$ B factories should make
both $\sin(2\beta)$ and $\Delta M_d$ precision observables, and the bounds
are expected to be improved by at least one order of magnitude, if we do
not discover SUSY by that time.

\vspace*{1cm}

\centerline{\bf Acknowledgement}

A.K. has been supported by the BRNS grant 2000/37/10/BRNS of DAE,
Govt.\ of India, by
the grant F.10-14/2001 (SR-I) of UGC, India, and
by the fellowship of the Alexander von Humboldt Foundation. J.P.S. thanks
CSIR, India, for a Research Fellowship.

\end{document}